\documentclass[aps,onecolumn,11pt,nofootinbib,tightenlines]{revtex4}
\usepackage[paper=a4paper,margin=24mm]{geometry}
\usepackage{graphicx,amsmath,amssymb,flafter,xcolor}
\usepackage[varg]{txfonts}
\usepackage{booktabs,siunitx}
\usepackage{placeins}
\usepackage[colorlinks=true,linkcolor=blue,citecolor=blue,urlcolor=blue]{hyperref}
\newcommand{\bx}{\mathbf e_x}
\newcommand{\by}{\mathbf e_y}
\newcommand{\bt}{\mathbf t}
\newcommand{\br}{\mathbf r}
\newcommand{\bb}{\mathbf b}
\newcommand{\order}{\mathcal O}
\begin{document}
\title{Beam Transport Through a Quadrupole with Offsets in Both Transverse Planes}
\author{Sergey Tomin}
\email{sergey.tomin@desy.de}
\affiliation{Deutsches Elektronen-Synchrotron DESY, Notkestrasse 85, 22607 Hamburg, Germany}
\date{\today}

\begin{abstract}
In accelerator switchyards with shared magnets, a beam may traverse a quadrupole with offsets in both transverse planes, a configuration that cannot be represented by a single conventional horizontal or vertical combined-function bend. Such an off-axis quadrupole bends the reference trajectory and generates dispersion simultaneously in both planes. We derive a compact analytical $6\times6$ transfer matrix about the curved reference trajectory. The model is formulated in Bishop coordinates using a weak-curvature approximation that retains the longitudinal variation of both reference-curvature components and the resulting dispersion in both transverse planes. The analytical linear model is compared with a sliced representation based on conventional second-order combined-function bend maps and with direct numerical integration of the Lorentz equations in the ideal Cartesian quadrupole field.
\end{abstract}

\maketitle

\section{Introduction}\label{sec:introduction}
\begin{figure}[!htbp]
\centering
\includegraphics[width=.64\textwidth]{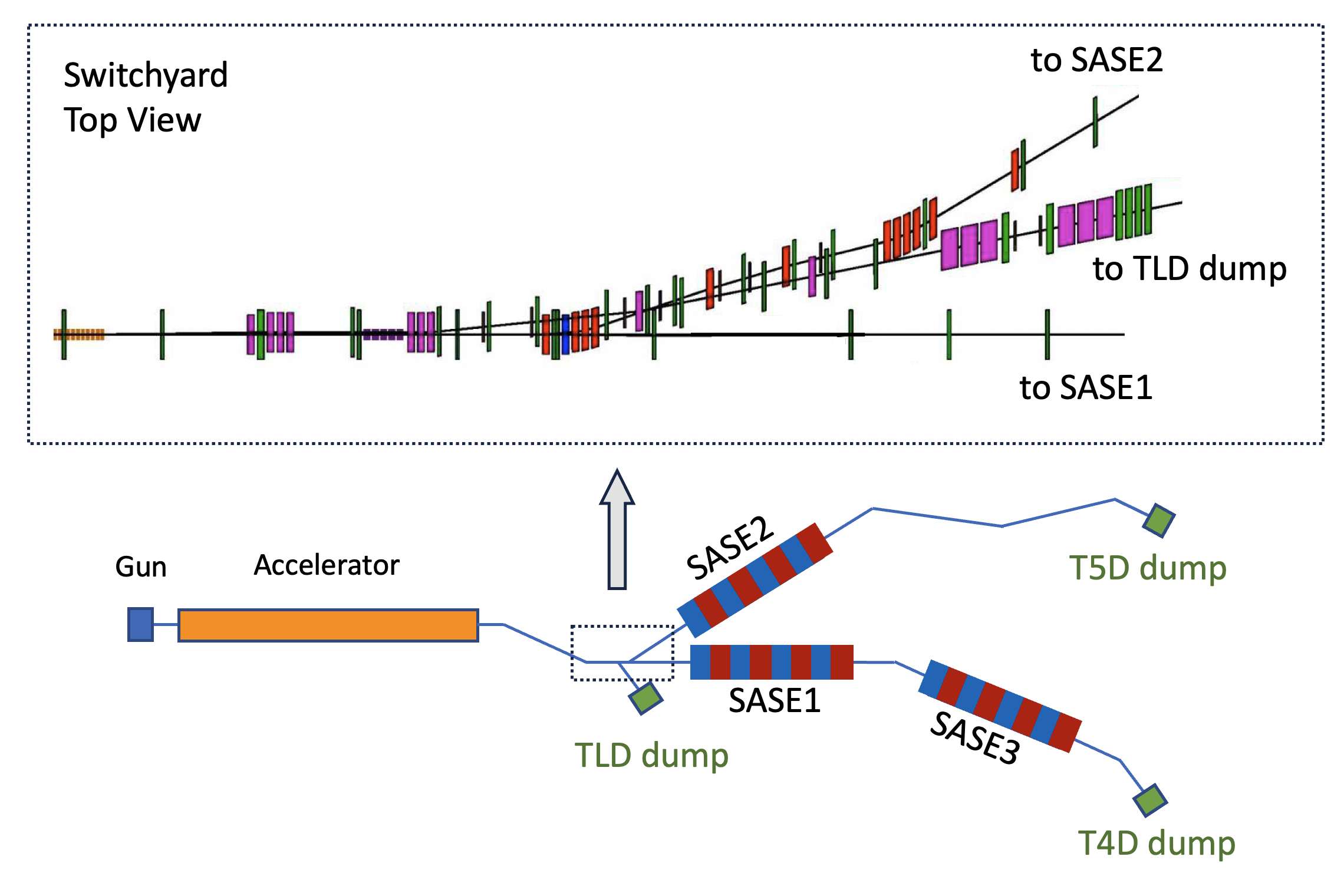}
\caption{Layout of the European XFEL switchyard and undulator branches.
Several magnets are shared between different beamlines.}
\label{fig:switch}
\end{figure}

The European XFEL superconducting accelerator was designed to produce electron
bunches with energies up to 17.5~GeV in burst-mode operation at a repetition
rate of 10~Hz. Each burst can contain up to 2700 electron bunches, yielding a
total of 27{,}000 bunches per second~\cite{EuXFEL}. Within each pulse train, a
fast kicker system in the switchyard distributes bunches towards three
undulator lines or directs them to the main accelerator dump if they are not
intended for lasing. The switchyard therefore constitutes a complex
beam-distribution system, see Fig.~\ref{fig:switch}.

Because of the high electron energy and the geometric constraints, it is
technically challenging to fully separate all beamlines. Consequently,
several quadrupole magnets in the switchyard effectively belong to multiple
lines: for one beamline a given magnet operates as a conventional quadrupole,
whereas for another beamline, passing through the same magnet with a transverse offset, it experiences both dipole bending and quadrupole focusing.
Even for an offset in only one transverse plane, an off-axis quadrupole
is not, in general, equivalent to a conventional combined-function
bend. As the reference particle moves through the quadrupole, its transverse
offset changes and hence the corresponding reference curvature varies along
the magnet. A conventional combined-function bend with prescribed constant
curvature can nevertheless provide a useful approximation in this case.

\begin{figure}[!htbp]
\centering
\includegraphics[width=.92\textwidth]{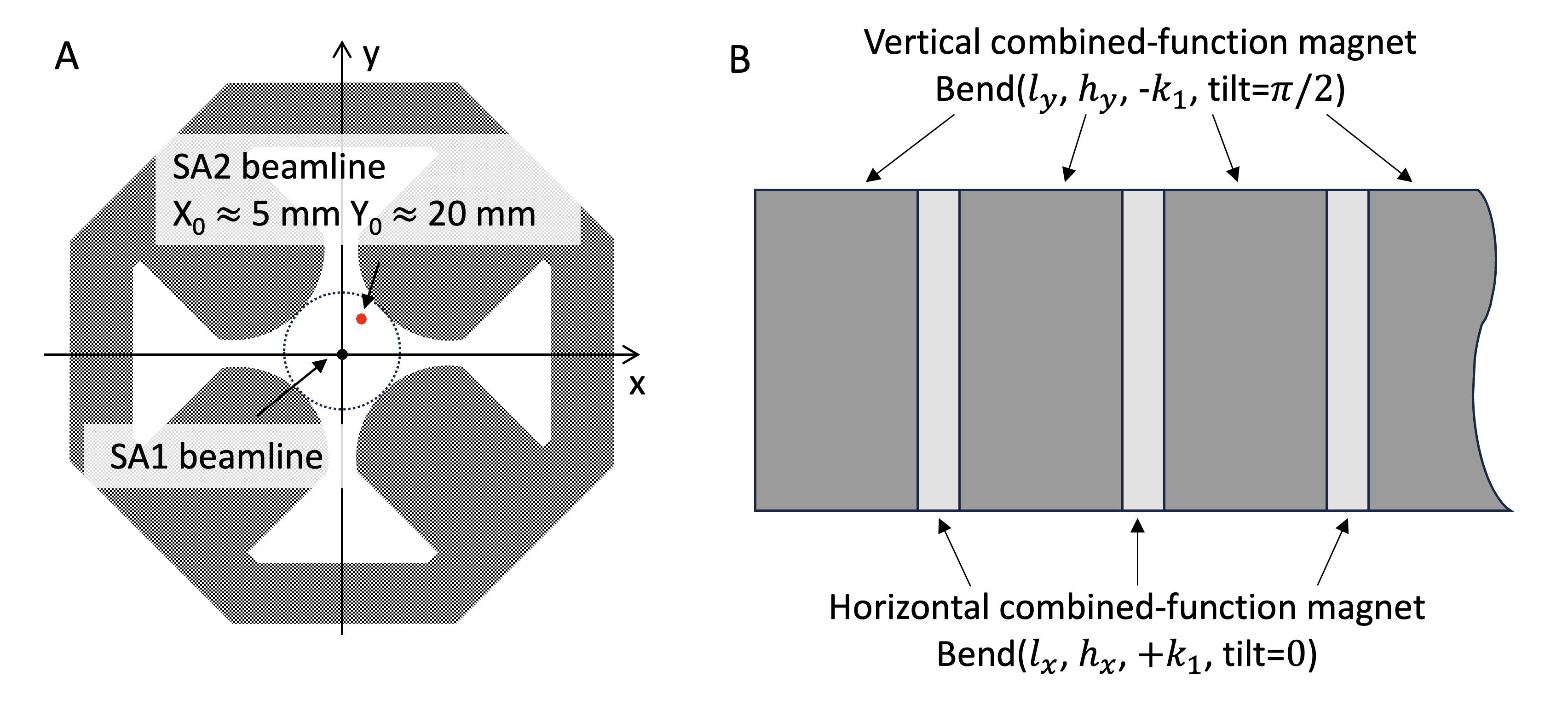}
\caption{(a) A reference entering a shared quadrupole with horizontal and
vertical offsets. (b) The original alternating horizontal and vertical
combined-function construction. The drawing is schematic.}
\label{fig:xyquad}
\end{figure}

The situation becomes more complicated when the beam has offsets in both
transverse planes, Fig.~\ref{fig:xyquad}(a). The reference trajectory then
bends simultaneously in both directions, with two generally varying curvature
components, and generates dispersion in both $x$ and $y$. Such a trajectory
cannot be represented by a single conventional horizontal or vertical
combined-function bend.

To address this limitation within existing accelerator codes, a sliced-bend representation has previously been implemented at the European XFEL, as illustrated in Fig.~\ref{fig:xyquad}(b). In this approach,
the quadrupole is decomposed into a sequence of alternating horizontal and
vertical combined-function segments. For parallel entry, the nominal
curvatures are set by the entrance offsets:
\[
h_x(0)=k_1a,\qquad h_y(0)=-k_1b,
\]
where $a=X_0(0)$ and $b=Y_0(0)$ are measured from the magnet axis, and
$k_1=qG/p_0$ is the normalized quadrupole strength, with signed charge $q$,
reference momentum $p_0$ and laboratory gradient $G=\partial B_Y/\partial X$.

 Numerical tracking showed that, for sufficiently short slices, this construction can provide a satisfactory approximation to the particle motion. Its applicability is expected when the curvature-related corrections remain small compared with the dominant quadrupole focusing. It is, however,  cumbersome for lattice modelling: a single physical magnet must be replaced by tens or hundreds of individual elements, making the lattice more complicated and the representation less convenient for codes such as MAD8 \cite{MAD8} or Ocelot \cite{ocelot}.

Instead of replacing the offset quadrupole by many short slices, we describe
it with a single linear matrix. This provides a convenient representation
within the \textsc{Ocelot} lattice structure, avoiding manual slicing of
long elements into hundreds of segments.

In this work, we derive an analytical linear model and the corresponding
$6\times6$ transfer matrix for a quadrupole traversed off axis in both
transverse planes. We first introduce the coordinate system and Hamiltonian
for the off-axis reference trajectory. We then derive the local vector
potential and, under the relevant small-curvature and small-angle
approximations, obtain the quadratic Hamiltonian and the corresponding
linear transfer matrix. The model is subsequently compared with numerical
tracking in the ideal quadrupole body field. The corresponding local vector
potential is derived in Appendix~\ref{app:gauge}. The magnetic model is a
straight, ideal constant-gradient body; fringe fields, radiation, and
collective effects are not included.

\section{Hamiltonian for a quadrupole with offsets in both planes}
\label{sec:hamiltonian}

\subsection{Coordinate system and Hamiltonian}

\begin{figure}[!htbp]
\centering
\includegraphics[width=.92\textwidth]{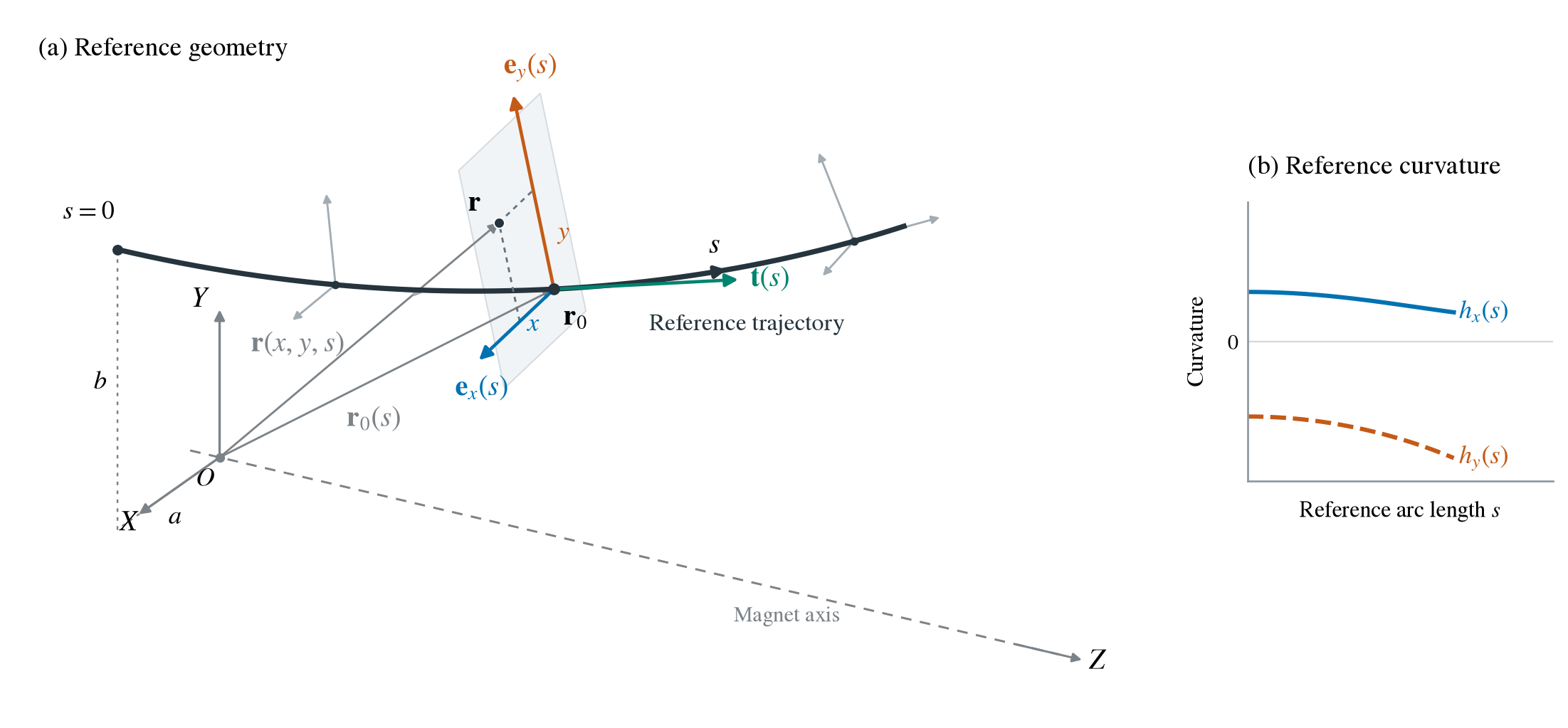}
\caption{Reference geometry for an off-axis trajectory
through an ideal quadrupole.
(a) The reference trajectory and associated Bishop frame.
(b) The corresponding curvature components along
the reference trajectory.}
\label{fig:bishop}
\end{figure}

For completeness, we briefly review the curvilinear coordinate
system and the magnetic Hamiltonian used below. The formulation
follows standard canonical beam-dynamics treatments~
\cite{Ripken95,Wolski,Iselin2000}. We employ a Bishop frame~
\cite{Bishop,Carroll} (see Fig.~\ref{fig:bishop}(a)), in which
the transverse basis is transported without rotation about the
reference tangent. The corresponding transport equations are
equivalent to those used by Ripken and Schmidt~\cite{Ripken95},
but we relax their restriction to piecewise-planar reference
trajectories, allowing simultaneous curvature in both transverse
directions.
We use capital letters $(X,Y,Z)$ for the fixed coordinates of
the magnet, with $Z$ along its axis. For an ideal quadrupole
with constant gradient $G=\partial B_Y/\partial X$, the field
and normalized strength are
\begin{equation}
\mathbf B
=
G\begin{pmatrix}
Y\\
X\\
0
\end{pmatrix},
\qquad
\bb
=
\frac{q\mathbf B}{p_0}
=
k_1
\begin{pmatrix}
Y\\
X\\
0
\end{pmatrix},
\qquad
k_1=\frac{qG}{p_0}.
\label{eq:labfield}
\end{equation}
Here $q$ is the signed particle charge and $p_0$ is the
reference momentum. With this convention, positive $k_1$
corresponds to horizontal focusing.

The reference trajectory $\br_0=(X_0,Y_0,Z_0)$ satisfies
\begin{equation}
\br_0'=\bt,\qquad
\bt'=\bt\times\bb(\br_0),\qquad
|\bt|=1,
\label{eq:reference_lorentz}
\end{equation}
where primes denote derivatives with respect to the reference
arc length $s$. Particle deviations from this trajectory are
described by the local transverse coordinates $x,y$.

Following the beam-optics notation of Ref.~\cite{Ripken95},
we write
\begin{equation}
\br(x,y,s)=\br_0(s)+x\bx(s)+y\by(s).
\label{eq:position}
\end{equation}
The right-handed orthonormal Bishop frame obeys
\begin{equation}
\bx'=h_x\bt,\qquad
\by'=h_y\bt,\qquad
\bt'=-h_x\bx-h_y\by.
\label{eq:bishop}
\end{equation}
Both curvature components $h_x$ and $h_y$ may be nonzero
simultaneously. The position differential is
\begin{equation}
d\br=\bx\,dx+\by\,dy+g\bt\,ds,
\qquad
g=1+h_xx+h_yy.
\label{eq:metric}
\end{equation}
Thus, the coordinates are orthogonal, with scale factors
$(1,1,g)$. We assume that particle deviations remain
sufficiently small for the local coordinates to be well
defined, with $g>0$.

We use the canonical variables
\begin{equation}
\mathbf z=(x,p_x,y,p_y,\sigma,\delta)^T,\qquad
\sigma=\frac{s}{\beta_0}-ct,\qquad
\delta=\frac{E-E_0}{p_0c}.
\label{eq:state}
\end{equation}
The reference particle has velocity $\beta_0c$, energy
$E_0=\gamma_0mc^2$, and momentum $p_0=\beta_0\gamma_0mc$.
Here $\delta$ is the normalized energy deviation used in
the canonical formulation~\cite{Wolski}, while a late
particle has negative $\sigma$.

Let $\mathbf A$ denote the magnetic vector potential,
$\mathbf B=\nabla\times\mathbf A$, and define its normalized
form $\mathbf a=q\mathbf A/p_0$. The transverse canonical
momenta $p_x,p_y$, normalized to $p_0$, are related to the
mechanical momenta by
\begin{equation}
\begin{aligned}
p_x&=\frac{\gamma m v_x+qA_x}{p_0},&
\frac{\pi_x}{p_0}&=p_x-a_x,\\
p_y&=\frac{\gamma m v_y+qA_y}{p_0},&
\frac{\pi_y}{p_0}&=p_y-a_y,
\end{aligned}
\label{eq:canonical_mechanical}
\end{equation}
where $\pi_i=\gamma m v_i$ and
$v_i=\mathbf v\cdot\mathbf e_i$ for $i\in\{x,y\}$.
Although the canonical momenta depend on the gauge,
the mechanical combinations $p_x-a_x$ and $p_y-a_y$
are gauge invariant.

In the curvilinear coordinates, the normalized vector potential satisfies
\begin{equation}
\mathbf a\cdot d\br
=
a_x\,dx+a_y\,dy+\mathcal A_s\,ds,
\qquad
\mathcal A_s=g\,a_s,\quad a_s=\mathbf a\cdot\bt.
\end{equation}
The corresponding standard Hamiltonian for forward
motion, with $s$ as the independent variable, is~
\cite{Ripken95,Wolski,Iselin2000}
\begin{equation}
H=\frac{\delta}{\beta_0}
-g\sqrt{
1+\frac{2\delta}{\beta_0}+\delta^2
-(p_x-a_x)^2-(p_y-a_y)^2
}
-\mathcal A_s.
\label{eq:exactH}
\end{equation}

Equation~\eqref{eq:exactH} is our starting point.
Since our objective is to obtain the analytical linear
transport matrix, we require the Hamiltonian only
through second order in the particle variables.
The construction of the local vector potential for
the ideal quadrupole body field and its reduction
to the form needed for the analytical linear model
are presented in Appendix~\ref{app:gauge}.

\subsection{Reference trajectory and geometrical ordering}

The curvatures \(h_x(s)\) and \(h_y(s)\) entering the curvilinear Hamiltonian
are determined by the reference trajectory through the quadrupole. We
therefore first obtain the reference orbit and its tangent to the accuracy
required for the analytical linear model, and then identify the geometrical
terms that can be neglected.

We restrict the analysis to paraxial reference trajectories that remain
weakly tilted with respect to the magnet axis throughout the quadrupole,
\begin{equation}
\theta_x^2(s)+\theta_y^2(s)\ll1,
\qquad
0\le s\le L,
\label{eq:weak_tilt}
\end{equation}
where
\[
\theta_x(s)\equiv t_X(s),
\qquad
\theta_y(s)\equiv t_Y(s)
\]
are the transverse components of the reference tangent.

We choose as reference trajectory the orbit of a particle entering the ideal
quadrupole with transverse offsets \(a\) and \(b\) and entrance angles
\[
\theta_{x0}\equiv\theta_x(0),
\qquad
\theta_{y0}\equiv\theta_y(0).
\]
The entrance conditions are therefore
\begin{equation}
X_0(0)=a,\qquad
Y_0(0)=b,\qquad
Z_0(0)=0,
\qquad
\bt(0)=
\begin{pmatrix}
\theta_{x0}\\[1mm]
\theta_{y0}\\[1mm]
\sqrt{1-\theta_{x0}^2-\theta_{y0}^2}
\end{pmatrix}
\simeq
\begin{pmatrix}
\theta_{x0}\\[1mm]
\theta_{y0}\\[1mm]
1
\end{pmatrix}.
\label{eq:entry}
\end{equation}
The last approximation follows directly from
Eq.~\eqref{eq:weak_tilt}.

In the paraxial approximation, the reference trajectory obeys the usual
quadrupole equations,
\begin{equation}
X_0''+k_1X_0=0,
\qquad
Y_0''-k_1Y_0=0,
\label{eq:reference_equations}
\end{equation}
where a prime denotes differentiation with respect to the reference arc
length \(s\). Since
\(t_Z=1+O(\theta_x^2+\theta_y^2)\), the reference arc length \(s\) and the
laboratory longitudinal coordinate \(Z\) differ only beyond the order
retained here.

The solution of Eq.~\eqref{eq:reference_equations} with the entrance
conditions of Eq.~\eqref{eq:entry} is
\begin{equation}
X_0(s)\simeq
aC_x(s)+\theta_{x0}S_x(s),
\qquad
Y_0(s)\simeq
bC_y(s)+\theta_{y0}S_y(s),
\label{eq:reference_orbit}
\end{equation}
where \(C_i\) and \(S_i\) are the usual quadrupole transport functions.
For \(k_1>0\),
\begin{equation}
\begin{aligned}
C_x(s)&=\cos(\sqrt{k_1}s),
&
S_x(s)&=\frac{\sin(\sqrt{k_1}s)}{\sqrt{k_1}},
\\
C_y(s)&=\cosh(\sqrt{k_1}s),
&
S_y(s)&=\frac{\sinh(\sqrt{k_1}s)}{\sqrt{k_1}}.
\end{aligned}
\label{eq:quadrupole_functions}
\end{equation}
They satisfy
\begin{equation}
C_x'=-k_1S_x,\qquad S_x'=C_x,
\qquad
C_y'=k_1S_y,\qquad S_y'=C_y.
\label{eq:quadrupole_identities}
\end{equation}
For \(k_1<0\), the focusing and defocusing functions are interchanged.

The transverse components of the reference tangent are consequently
\begin{equation}
\begin{aligned}
\theta_x(s)\equiv t_X(s)
&\simeq
-k_1aS_x(s)+\theta_{x0}C_x(s),
\\
\theta_y(s)\equiv t_Y(s)
&\simeq
k_1bS_y(s)+\theta_{y0}C_y(s).
\end{aligned}
\label{eq:reference_angles}
\end{equation}

The curvature of the reference trajectory follows directly from the Lorentz
equation. Comparing
\[
\bt'=\bt\times\bb(\br_0)
\]
with the Bishop-frame relation
\[
\bt'=-h_x\bx-h_y\by
\]
gives
\begin{equation}
h_x=\bb(\br_0)\cdot\by,
\qquad
h_y=-\bb(\br_0)\cdot\bx.
\label{eq:curvatures}
\end{equation}

Under the small-angle condition of Eq.~\eqref{eq:weak_tilt}, the local
transverse basis remains aligned with the corresponding laboratory
transverse directions to the accuracy required here. Substituting the ideal
quadrupole field of Eq.~\eqref{eq:labfield} into
Eq.~\eqref{eq:curvatures} therefore gives
\begin{equation}
\begin{aligned}
h_x(s)
&\simeq k_1X_0(s)
\simeq
k_1\left[
aC_x(s)+\theta_{x0}S_x(s)
\right],
\\
h_y(s)
&\simeq -k_1Y_0(s)
\simeq
-k_1\left[
bC_y(s)+\theta_{y0}S_y(s)
\right].
\end{aligned}
\label{eq:lin_reference}
\end{equation}

Thus, unlike in a conventional combined-function bend with prescribed constant curvature, the curvature generated by an off-axis quadrupole generally varies along the reference trajectory, as illustrated schematically in Fig.~\ref{fig:bishop}(b).

In addition to the paraxial condition of Eq.~\eqref{eq:weak_tilt},
the analytical model neglects curvature-induced focusing terms
that are quadratic in the reference curvatures. We therefore
require
\begin{equation}
h_x^2(s),\,h_y^2(s),\,|h_x(s)h_y(s)|
\ll |k_1|.
\label{eq:weak_curvature}
\end{equation}
Using $h_x\simeq k_1X_0$ and $h_y\simeq -k_1Y_0$,
this condition can be expressed as
\begin{equation}
|k_1|X_0^2(s)\ll 1,\qquad
|k_1|Y_0^2(s)\ll 1,
\qquad 0\le s\le L.
\label{eq:weak_offsets}
\end{equation}
Thus, in addition to the small reference angles required by
Eq.~\eqref{eq:weak_tilt}, the reference offsets must remain
small compared with the characteristic quadrupole scale
$1/\sqrt{|k_1|}$.

\subsection{Vector potential for the analytical linear model}

The vector potential used in the analytical linear model has the same bending-plus-quadrupole structure as that employed in Ref.~\cite{Ripken95}. Here, we briefly outline its construction from the ideal Cartesian quadrupole field in Bishop coordinates and identify the approximations under which this form is recovered for a reference trajectory with simultaneous, longitudinally varying curvature in both transverse planes.

We now construct the local vector potential required for the quadratic
Hamiltonian. On the reference trajectory, Eq.~\eqref{eq:curvatures} gives
\[
b_x=-h_y,
\qquad
b_y=h_x.
\]
To first order in the transverse displacement from the reference trajectory,
the quadrupole field therefore becomes
\begin{equation}
b_x=-h_y+k_1y,
\qquad
b_y=h_x+k_1x.
\label{eq:lin_transverse_field}
\end{equation}

The longitudinal field component is of higher reference order. On the
reference trajectory,
\begin{equation}
b_s=\bb(\br_0)\cdot\bt
=k_1\left(Y_0\theta_x+X_0\theta_y\right),
\end{equation}
and is therefore second order in the reference offsets and angles. Its
variation across the transverse plane first contributes beyond the quadratic
Hamiltonian retained here. We therefore use
\begin{equation}
b_x=-h_y+k_1y,
\qquad
b_y=h_x+k_1x,
\qquad
b_s=0.
\label{eq:lin_local_field}
\end{equation}

At this order a gauge may be chosen with
\[
a_x=a_y=0.
\]
The Bishop-coordinate curl equations then reduce to
\begin{equation}
b_x=\frac{1}{g}\partial_y\mathcal A_s,
\qquad
b_y=-\frac{1}{g}\partial_x\mathcal A_s,
\label{eq:lin_curl}
\end{equation}
where
\[
g=1+h_xx+h_yy.
\]
Terms generated by the factor \(g\) that are quadratic in the reference
curvatures are neglected according to Eq.~\eqref{eq:weak_curvature}, while
terms quadratic in the transverse coordinates contribute only beyond the
quadratic Hamiltonian. To the required accuracy,
\[
\partial_y\mathcal A_s=-h_y+k_1y,
\qquad
\partial_x\mathcal A_s=-h_x-k_1x,
\]
and hence
\begin{equation}
a_x=a_y=0,
\qquad
\mathcal A_s
=a_0(s)-h_x(s)x-h_y(s)y-\frac{k_1}{2}(x^2-y^2).
\label{eq:lin_potential}
\end{equation}
Here \(a_0(s)=\mathcal A_s(0,0,s)\) is gauge dependent and does not affect
the equations of motion. 

The resulting expression agrees, to the order retained, with the conventional bending-plus-quadrupole vector potential of Ref.~\cite{Ripken95}. The full construction and the higher-order terms neglected in this approximation are discussed in Appendix~\ref{app:gauge}.

\subsection{Quadratic Hamiltonian}

We now expand the Hamiltonian in the particle variables, counting
\(x,p_x,y,p_y,\delta\) as first degree. The reference quantities
\(h_x(s)\) and \(h_y(s)\) are treated as prescribed coefficient functions.
With
\[
u=h_xx+h_yy,
\]
substitution of Eq.~\eqref{eq:lin_potential} into
Eq.~\eqref{eq:exactH} gives, through second particle degree,
\begin{equation}
\begin{aligned}
\sqrt{1+2\delta/\beta_0+\delta^2-p_x^2-p_y^2}
&=
1+\frac{\delta}{\beta_0}
-\frac{p_x^2+p_y^2}{2}
-\frac{\delta^2}{2\beta_0^2\gamma_0^2}
+\order(3),
\\
g\sqrt{\cdots}
&=
1+u+\frac{\delta}{\beta_0}
+\frac{u\delta}{\beta_0}
-\frac{p_x^2+p_y^2}{2}
-\frac{\delta^2}{2\beta_0^2\gamma_0^2}
+\order(3).
\end{aligned}
\label{eq:sqrt_expansion}
\end{equation}

The term linear in \(\delta\) cancels between the first term of
Eq.~\eqref{eq:exactH} and the square-root contribution. Likewise, the term
\(-u\) from the square-root part cancels the \(+u\) contribution from
\(-\mathcal A_s\). Thus the terms linear in the particle variables vanish,
as expected for an expansion about the reference trajectory. The remaining
term \(-1-a_0(s)\), which is independent of the phase-space variables, may
be dropped without affecting the equations of motion.

The quadratic Hamiltonian of the analytical linear model is therefore
\begin{equation}
\boxed{
H_{2,\mathrm{lin}}=
\frac{p_x^2+p_y^2}{2}
+\frac{k_1}{2}(x^2-y^2)
-\frac{\delta}{\beta_0}
\left[h_x(s)x+h_y(s)y\right]
+\frac{\delta^2}{2\beta_0^2\gamma_0^2}.
}
\label{eq:hlin}
\end{equation}

The term
\[
-\frac{\delta}{\beta_0}
\left[h_x(s)x+h_y(s)y\right]
\]
couples the energy deviation to the transverse motion and drives dispersion
in both planes.

\section{Analytical linear transfer matrix}
\label{sec:matrix}

\subsection{Hamilton equations}

We now solve the linear equations generated by the quadratic Hamiltonian
of Eq.~\eqref{eq:hlin}. For the canonical pairs
\((x,p_x)\), \((y,p_y)\), and \((\sigma,\delta)\), Hamilton's equations give
\begin{equation}
\begin{aligned}
x'
&=\frac{\partial H_{2,\mathrm{lin}}}{\partial p_x}
=p_x,
&
p_x'
&=-\frac{\partial H_{2,\mathrm{lin}}}{\partial x}
=-k_1x+\frac{h_x(s)}{\beta_0}\delta,
\\
y'
&=\frac{\partial H_{2,\mathrm{lin}}}{\partial p_y}
=p_y,
&
p_y'
&=-\frac{\partial H_{2,\mathrm{lin}}}{\partial y}
=k_1y+\frac{h_y(s)}{\beta_0}\delta,
\\
\sigma'
&=\frac{\partial H_{2,\mathrm{lin}}}{\partial\delta}
=\frac{\delta}{\beta_0^2\gamma_0^2}
-\frac{h_x(s)x+h_y(s)y}{\beta_0},
&
\delta'
&=-\frac{\partial H_{2,\mathrm{lin}}}{\partial\sigma}
=0.
\end{aligned}
\label{eq:eom}
\end{equation}
In particular, \(\delta\) is constant through the element. Combining the
canonical equations in each transverse plane gives
\begin{equation}
x''+k_1x=\frac{h_x(s)}{\beta_0}\delta,
\qquad
y''-k_1y=\frac{h_y(s)}{\beta_0}\delta.
\label{eq:forced_transverse}
\end{equation}

\subsection{Homogeneous quadrupole motion}

Without the energy-dependent forcing, Eq.~\eqref{eq:forced_transverse}
reduces to the ordinary quadrupole equations already encountered for the
reference trajectory. Using the same transport functions \(C_i\) and \(S_i\),
the homogeneous solution is therefore
\begin{equation}
x_{\mathrm h}(s)
=C_x(s)x(0)+S_x(s)p_x(0),
\qquad
y_{\mathrm h}(s)
=C_y(s)y(0)+S_y(s)p_y(0).
\label{eq:homogeneous_solution}
\end{equation}
Here \(x'=p_x\) and \(y'=p_y\) follow directly from Eq.~\eqref{eq:eom}.

\subsection{Dispersive contribution}

Since \(\delta\) is constant, the full transverse solution can be written as
the homogeneous quadrupole motion plus the response to the energy deviation,
\begin{equation}
\begin{aligned}
x(s)
&=C_x(s)x(0)+S_x(s)p_x(0)+D_x(s)\delta(0),
\\
y(s)
&=C_y(s)y(0)+S_y(s)p_y(0)+D_y(s)\delta(0).
\end{aligned}
\label{eq:transverse_solution}
\end{equation}
Substituting Eq.~\eqref{eq:transverse_solution} into
Eq.~\eqref{eq:forced_transverse} gives
\begin{equation}
D_x''+k_1D_x=\frac{h_x(s)}{\beta_0},
\qquad
D_y''-k_1D_y=\frac{h_y(s)}{\beta_0},
\label{eq:dispersion_eq_general}
\end{equation}
with
\[
D_x(0)=D_x'(0)=D_y(0)=D_y'(0)=0.
\]

Using the varying reference curvatures of
Eq.~\eqref{eq:lin_reference},
\[
h_x(s)\simeq
k_1\left[aC_x(s)+\theta_{x0}S_x(s)\right],
\qquad
h_y(s)\simeq
-k_1\left[bC_y(s)+\theta_{y0}S_y(s)\right],
\]
the dispersion equations become
\begin{equation}
\begin{aligned}
D_x''+k_1D_x
&=
\frac{k_1}{\beta_0}
\left[aC_x(s)+\theta_{x0}S_x(s)\right],
\\
D_y''-k_1D_y
&=
-\frac{k_1}{\beta_0}
\left[bC_y(s)+\theta_{y0}S_y(s)\right].
\end{aligned}
\label{eq:dispersion_eq}
\end{equation}

The particular \(s\) dependence of the reference curvatures again allows
the dispersion equations to be solved analytically. In addition to
\begin{equation}
\left(\frac{sS_x}{2}\right)''
+k_1\frac{sS_x}{2}=C_x,
\qquad
\left(\frac{sS_y}{2}\right)''
-k_1\frac{sS_y}{2}=C_y,
\label{eq:dispersion_identity}
\end{equation}
the sine-like terms satisfy
\begin{equation}
\left(\frac{S_x-sC_x}{2k_1}\right)''
+k_1\frac{S_x-sC_x}{2k_1}=S_x,
\qquad
\left(\frac{sC_y-S_y}{2k_1}\right)''
-k_1\frac{sC_y-S_y}{2k_1}=S_y.
\label{eq:dispersion_identity_angle}
\end{equation}
These identities provide the particular solutions associated with the \(C_i(s)\)- and \(S_i(s)\)-dependent forcing terms in Eq.~\eqref{eq:dispersion_eq}.
We therefore obtain 
\begin{equation}
\begin{aligned}
D_x(s)
&=
\frac{k_1a}{2\beta_0}sS_x(s)
+\frac{\theta_{x0}}{2\beta_0}
\left[S_x(s)-sC_x(s)\right],
\\
D_y(s)
&=
-\frac{k_1b}{2\beta_0}sS_y(s)
+\frac{\theta_{y0}}{2\beta_0}
\left[S_y(s)-sC_y(s)\right].
\end{aligned}
\label{eq:dispersion_functions}
\end{equation}

At the exit reference arc length \(s=L\),
\begin{equation}
\begin{aligned}
D_x
&=
\frac{1}{2\beta_0}
\left[
k_1aLS_x
+\theta_{x0}(S_x-LC_x)
\right],
\\
D_x'
&=
\frac{k_1}{2\beta_0}
\left[
a(S_x+LC_x)
+\theta_{x0}LS_x
\right],
\\[1mm]
D_y
&=
\frac{1}{2\beta_0}
\left[
-k_1bLS_y
+\theta_{y0}(S_y-LC_y)
\right],
\\
D_y'
&=
-\frac{k_1}{2\beta_0}
\left[
b(S_y+LC_y)
+\theta_{y0}LS_y
\right].
\end{aligned}
\label{eq:dispersion}
\end{equation}
Here and below, unlabelled functions \(C_i\) and \(S_i\) are evaluated at
\(s=L\).

\subsection{Longitudinal motion and transfer matrix}

The longitudinal equation in Eq.~\eqref{eq:eom} is
\begin{equation}
\sigma'
=
\frac{\delta}{\beta_0^2\gamma_0^2}
-\frac{h_x(s)x+h_y(s)y}{\beta_0}.
\label{eq:sigma_eq}
\end{equation}
Since \(\delta\) is constant, the transverse solutions obtained above
determine the complete longitudinal response.

Because the map is generated by the quadratic Hamiltonian
Eq.~\eqref{eq:hlin}, it is symplectic. Once the dispersive coefficients
\(D_x,D_x',D_y,D_y'\) are known, symplecticity fixes the corresponding
transverse-to-longitudinal coefficients:
\begin{equation}
\begin{aligned}
U_x&=-k_1S_xD_x-C_xD_x',
&
V_x&=C_xD_x-S_xD_x',
\\
U_y&= k_1S_yD_y-C_yD_y',
&
V_y&=C_yD_y-S_yD_y'.
\end{aligned}
\label{eq:timing_symplectic}
\end{equation}

The remaining longitudinal coefficient follows directly from integrating
Eq.~\eqref{eq:sigma_eq},
\begin{equation}
R_{56}
=
\frac{L}{\beta_0^2\gamma_0^2}
-\frac{1}{\beta_0}
\int_0^L
\left[
h_x(s)D_x(s)+h_y(s)D_y(s)
\right]ds .
\label{eq:r56_integral}
\end{equation}

Collecting the transverse and longitudinal solutions, the transfer matrix
in the coordinate order of Eq.~\eqref{eq:state} is therefore
\begin{equation}
\boxed{
R_{\mathrm{lin}}(L)=
\begin{pmatrix}
C_x&S_x&0&0&0&D_x\\
-k_1S_x&C_x&0&0&0&D_x'\\
0&0&C_y&S_y&0&D_y\\
0&0&k_1S_y&C_y&0&D_y'\\
U_x&V_x&U_y&V_y&1&R_{56}\\
0&0&0&0&0&1
\end{pmatrix}.
}
\label{eq:R}
\end{equation}
All transverse and dispersive coefficients are given explicitly above.
Equation~\eqref{eq:R} is symplectic in the canonical variables of
Eq.~\eqref{eq:state}.

For completeness, the integral in Eq.~\eqref{eq:r56_integral} can also be
evaluated analytically. The result is
\begin{equation}
R_{56}
=
\frac{L}{\beta_0^2\gamma_0^2}
-\frac{\mathcal F_x+\mathcal F_y}{8\beta_0^2},
\label{eq:r56}
\end{equation}
where the explicit expressions for \(\mathcal F_x\) and \(\mathcal F_y\)
are given in Appendix~\ref{app:r56}.

\subsection{One-plane limit and constant-curvature approximation}

Consider the special case of a horizontal entrance offset with no vertical
offset or angle and, for simplicity, zero horizontal entrance angle,
\[
b=\theta_{y0}=\theta_{x0}=0.
\]
The reference curvature then has only a horizontal component,
\begin{equation}
h_x(s)\simeq k_1aC_x(s),
\qquad
h_y(s)=0.
\label{eq:oneplane_curvature}
\end{equation}
Thus an off-axis quadrupole in one transverse plane resembles a
combined-function magnet, but its effective curvature is generally not
constant along the reference trajectory.

A conventional combined-function element with the same entrance curvature
would instead use
\[
h_0=k_1a=\mathrm{const}.
\]
Within the weak-curvature approximation, the curvature-induced focusing
term \(h_0^2\) is negligible compared with \(k_1\), since
\[
\frac{h_0^2}{|k_1|}=|k_1|a^2\ll1.
\]
The principal difference between the two descriptions therefore comes from
the longitudinal variation of the reference curvature. For example, the
horizontal position-dispersion coefficients are
\begin{equation}
D_x=
\frac{k_1aLS_x}{2\beta_0},
\qquad
D_x^{(h_0)}
=
\frac{a}{\beta_0}(1-C_x),
\label{eq:constant_curvature_comparison}
\end{equation}
for the offset quadrupole and the constant-curvature approximation,
respectively. The two coincide in the weak-lens limit
\(|k_1|L^2\ll1\), but need not agree for a sufficiently strong or long
quadrupole.

\section{Comparison with numerical tracking and the bend-slice model}
\label{sec:comparison}

We compare the analytical linear model of Eq.~\eqref{eq:R} with two
independent calculations. The first is direct numerical integration of the
Lorentz equations in the ideal Cartesian quadrupole field and is used as the
reference solution. The second is a conventional bend-slice representation,
in which the off-axis quadrupole is approximated by a sequence of horizontal
and vertical combined-function bends, see Fig.~\ref{fig:xyquad}.

For the illustrative example considered below we use
\begin{equation}
    L=1~\mathrm{m},
    \qquad
    k_1=1~\mathrm{m}^{-2},
    \qquad
    (a,b)=(0.02,0.03)~\mathrm{m},
    \qquad
    (\theta_{y0}=\theta_{x0}=0)~\mathrm{rad},
    \label{eq:comparison_parameters}
\end{equation}
with parallel entrance conditions and $\beta_0=1$. \(L\) denotes the reference arc length

\subsection{Bend-slice model}

For an entrance offset $(a,b)$, the nominal horizontal and vertical bending
components are fixed by
\begin{equation}
    h_x=k_1 a,
    \qquad
    h_y=-k_1 b .
\end{equation}
The quadrupole is divided into $n$ longitudinal intervals. Each interval is
represented by a horizontal and a vertical combined-function bend of length
$L/(2n)$. Since each bending component acts over only half of the interval,
the corresponding bending angles are
\begin{equation}
    \theta_x=\frac{h_xL}{n},
    \qquad
    \theta_y=\frac{h_yL}{n}.
\end{equation}
Such a representation can be constructed using standard combined-function
bend elements available in accelerator codes such as
MAD8/MAD-X and \textsc{Ocelot}~\cite{MAD8,MADX,ocelot}.
The calculations shown here use the second-order maps implemented in
\textsc{Ocelot}. We use $n=8192$, for which further refinement produces
negligible changes in the quantities shown below.

\subsection{Direct numerical reference}

As an independent reference, particle trajectories are obtained by direct
integration of the Lorentz equations in the ideal quadrupole field
of Eq.~\eqref{eq:labfield}. Using the laboratory coordinate $Z$ as the
independent variable and normalized mechanical momenta
$P_i=p_i/p_0$, the transverse equations are
\begin{equation}
\begin{aligned}
    \frac{dX}{dZ} &= \frac{P_X}{P_Z},
    &\qquad
    \frac{dP_X}{dZ} &= -k_1X,\\
    \frac{dY}{dZ} &= \frac{P_Y}{P_Z},
    &
    \frac{dP_Y}{dZ} &= k_1Y ,
\end{aligned}
\label{eq:comparison_cartesian_tracking}
\end{equation}
with
\begin{equation}
    P_Z=
    \sqrt{
        1+\frac{2\delta}{\beta_0}+\delta^2
        -P_X^2-P_Y^2
    } .
    \label{eq:comparison_longitudinal_momentum}
\end{equation}
The trajectories are integrated with the DOP853 algorithm. The numerical
solution therefore uses only the Cartesian quadrupole field and does not rely
on either the curvilinear Hamiltonian or the bend-slice approximation.

To compare the numerical trajectories with the analytical transfer matrix, the particle coordinates and momenta must be expressed in the same curvilinear coordinate system. Both the reference particle and the test particles are tracked in the fixed Cartesian coordinates of the magnet. At a given reference arc length $s$, we identify the point $\mathbf r$ on the test-particle trajectory that lies in the plane perpendicular to the reference tangent $\mathbf t(s)$. This point is determined by the condition
\begin{equation}
\left[\mathbf r-\mathbf r_0(s)\right]\cdot\mathbf t(s)=0.
\end{equation}
The displacement from the reference trajectory, $\Delta\mathbf r=\mathbf r-\mathbf r_0(s)$, is then projected onto the local Bishop basis to obtain the transverse coordinates,
\begin{equation}
x=\Delta\mathbf r\cdot\mathbf e_x(s),
\qquad
y=\Delta\mathbf r\cdot\mathbf e_y(s).
\label{eq:numerical_coordinates}
\end{equation}
The corresponding transverse mechanical momenta are obtained by projecting the Cartesian particle momentum $\mathbf P$ onto the same basis,
\begin{equation}
\frac{\pi_x}{p_0}=\mathbf P\cdot\mathbf e_x(s),
\qquad
\frac{\pi_y}{p_0}=\mathbf P\cdot\mathbf e_y(s).
\label{eq:numerical_momenta}
\end{equation}
The projected mechanical momenta are converted to canonical
momenta according to
\[
p_x = \frac{\pi_x}{p_0} + a_x,
\qquad
p_y = \frac{\pi_y}{p_0} + a_y,
\]
where the transverse vector-potential components are
evaluated in the Poincar\'e gauge of Appendix~A,
using Eq.~(A8). The inverse transformation is applied
when initializing particles from canonical phase-space
coordinates. Although these components are neglected
in the analytical weak-curvature model, they are
retained in the numerical coordinate transformation.

The longitudinal coordinate is evaluated using the same convention as in
Eq.~\eqref{eq:state},
\begin{equation}
\sigma=\frac{s}{\beta_0}-ct,
\end{equation}
with the particle arrival time obtained together with the numerical
trajectory. The resulting numerical state is therefore expressed in the
same coordinates
\begin{equation}
\boldsymbol{\xi}
=
(x,p_x,y,p_y,\sigma,\delta)^T
\end{equation}
as the analytical map. The numerical derivatives below are evaluated about
the reference state \(\boldsymbol{\xi}=0\).

For each initial coordinate \(\xi_j\), we introduce a small perturbation \(\epsilon_j\), while keeping all other initial coordinates unchanged. Two particles are therefore tracked, with
\[
\xi_j=+\epsilon_j
\qquad\text{and}\qquad
\xi_j=-\epsilon_j.
\]

Let \(\xi_i^{f}(s;+\epsilon_j)\) and \(\xi_i^{f}(s;-\epsilon_j)\) denote the corresponding values of the \(i\)-th phase-space coordinate after tracking to position \(s\). The centered finite-difference derivative is then

\begin{equation}
D_{ij}(s;\epsilon_j)
=
\frac{
\xi_i^{f}(s;+\epsilon_j)
-
\xi_i^{f}(s;-\epsilon_j)
}{
2\epsilon_j
},
\qquad i,j=1,\ldots,6.
\label{eq:numerical_jacobian}
\end{equation}

For example, the element \(R_{11}\), which describes the dependence of the final horizontal position on the initial horizontal position, is evaluated as

\begin{equation}
D_{11}(s;\epsilon_x)
=
\frac{
x_f(s;x_i=+\epsilon_x)
-
x_f(s;x_i=-\epsilon_x)
}{
2\epsilon_x
},
\end{equation}

with all other initial phase-space coordinates kept fixed. The remaining matrix elements are obtained in the same way by perturbing each initial coordinate in turn.

For sufficiently small \(\epsilon_j\), the centered finite difference has the expansion

\begin{equation}
D_{ij}(s;\epsilon_j)
=
R_{ij}(s)
+
c_{ij}(s)\epsilon_j^2
+
\mathcal O(\epsilon_j^4),
\end{equation}

where \(R_{ij}(s)\) is the corresponding element of the first-order transfer matrix.

To reduce the dependence on the chosen perturbation size, the calculation is repeated with the step \(\epsilon_j/2\). The leading \(\mathcal O(\epsilon_j^2)\) finite-difference error is then removed using Richardson extrapolation~\cite{Burden2016},

\begin{equation}
R_{ij}^{\mathrm{num}}(s)
=
\frac{
4D_{ij}(s;\epsilon_j/2)
-
D_{ij}(s;\epsilon_j)
}{3}
+
\mathcal O(\epsilon_j^4).
\label{eq:numerical_R_matrix}
\end{equation}

Repeating this procedure for all six initial phase-space coordinates gives the complete numerical \(6\times6\) first-order transfer matrix \(R^{\mathrm{num}}(s)\).

\subsection{Comparison}

\begin{figure}[!htbp]
    \centering
    \includegraphics[width=\textwidth]
    {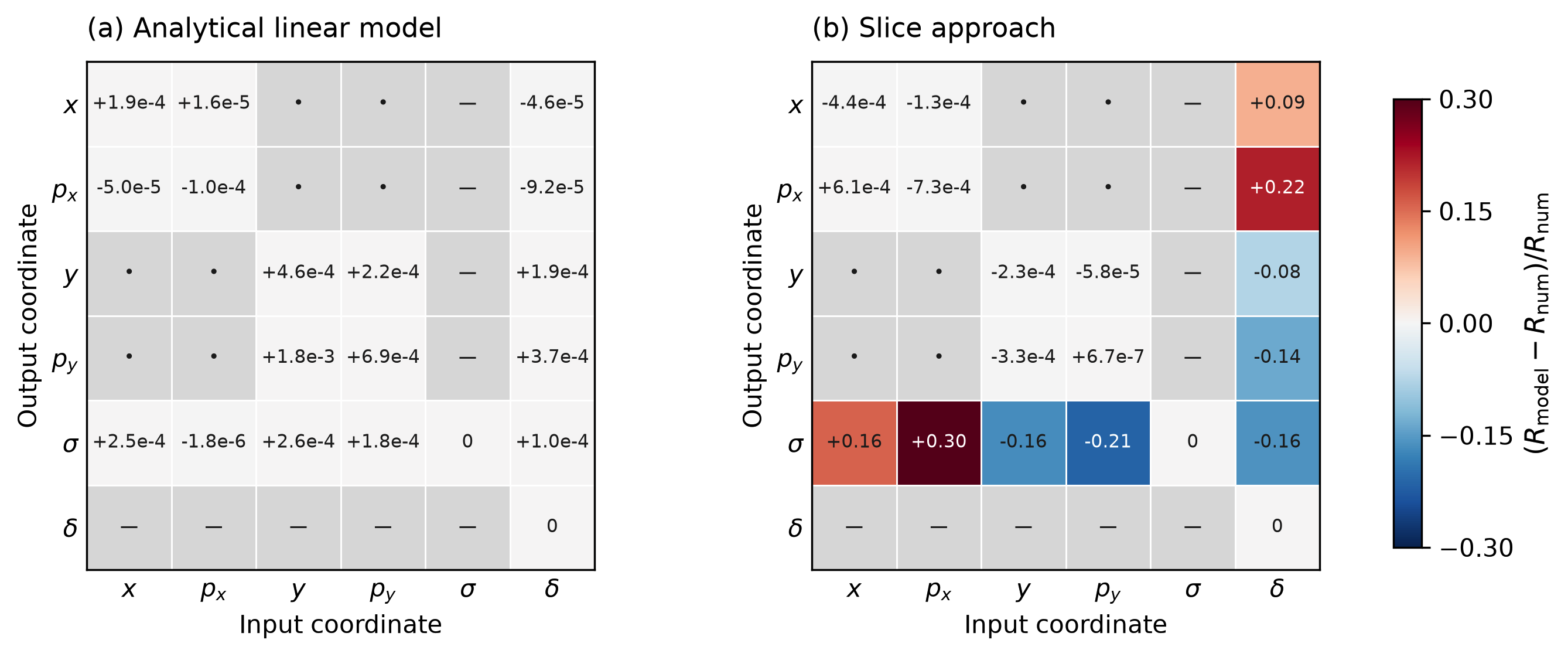}
    \caption{
    Comparison of the first-order transfer matrices
    in canonical coordinates with direct numerical tracking.
    (a) Analytical linear model.
    (b) Bend-slice model.
    Gray dashes indicate elements below numerical resolution,
    while gray dots indicate excluded cross-plane coupling
    terms. Both panels use the same relative-error scale.
    }
    \label{fig:Rmatrix_comparison}
\end{figure}

Figure~\ref{fig:Rmatrix_comparison} compares the first-order
transfer matrices of the analytical linear model (ALM)
and the bend-slice model with direct numerical tracking.
For the matrix elements included in the comparison, we
show the relative difference
\begin{equation}
    \Delta_{ij}^{(m)}
    =
    \frac{R_{ij}^{(m)}-R_{ij}^{\mathrm{num}}}
    {R_{ij}^{\mathrm{num}}},
    \qquad m\in\{\mathrm{ALM},\mathrm{slice}\}.
    \label{eq:relative_error}
\end{equation}
Elements below the numerical resolution and transverse
cross-plane coupling terms, which are neglected in both
approximate models, are excluded from the comparison.
For the sliced model, the matrix coefficients are taken
in its native bend-map coordinates; the reference
geometry is examined separately below.

Both models show comparable accuracy for the transverse
betatron motion, with the bend-slice model being slightly
more accurate for some elements due to the retained
curvature-squared focusing terms. A more significant
difference is found in the dispersive and longitudinal
coefficients, where the analytical model agrees
substantially better with numerical tracking.
This improvement results from retaining the longitudinal
variation of the reference curvature.

We next compare the transverse particle deviations.
The analytical model uses the linear matrix of
Eq.~\eqref{eq:R}, whereas the bend-slice model uses the
full second-order maps of the individual slices.
The error relative to direct numerical tracking is
defined as
\begin{equation}
    \Delta r_\perp
    =
    \sqrt{
        (x-x_{\mathrm{num}})^2
        +
        (y-y_{\mathrm{num}})^2
    },
    \label{eq:comparison_particle_error}
\end{equation}
where the transverse coordinates are measured relative
to the reference trajectory of each model at $s=L$.
Thus, the comparison concerns local particle transport
and does not include errors in the reference survey.

\begin{figure}[!htbp]
    \centering
    \includegraphics[width=\textwidth]
    {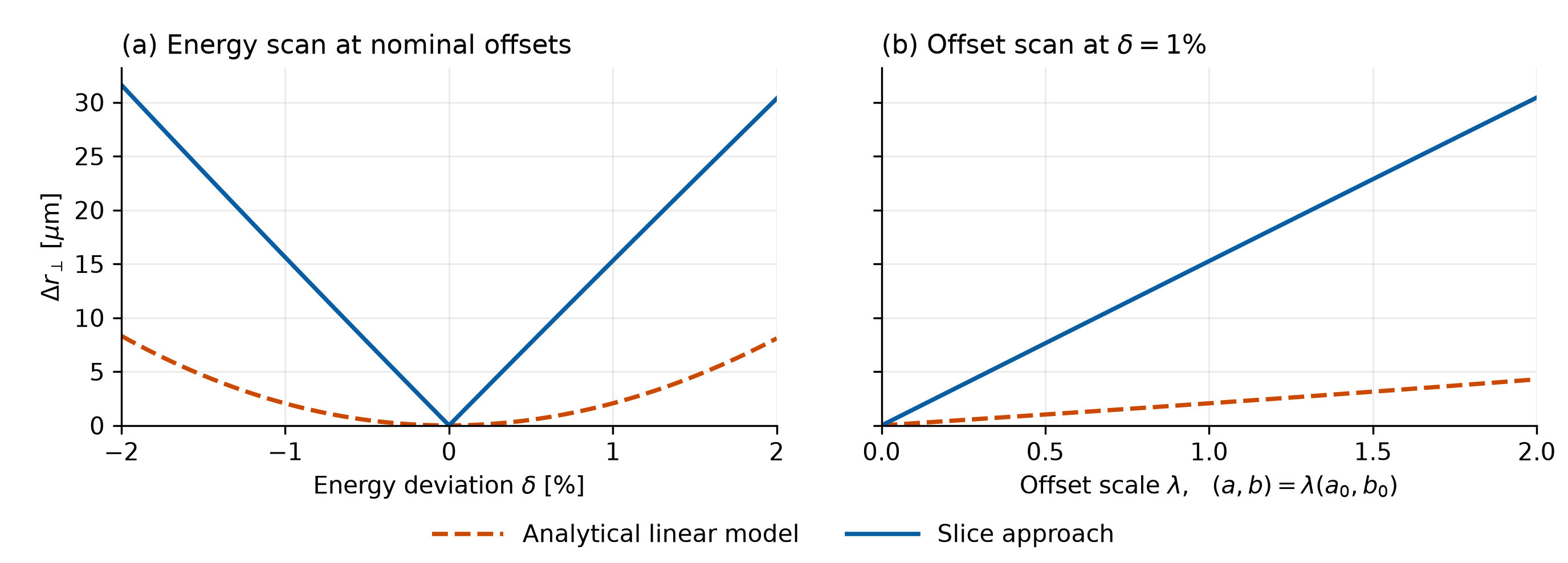}
    \caption{
    Comparison of transverse particle deviations
    relative to the reference trajectory.
    (a) Exit error as a function of the entrance
    energy deviation.
    (b) Exit error as a function of the offset scale
    at $\delta=1\%$.
    Dashed and solid curves represent the analytical
    and bend-slice models, respectively.
    }
    \label{fig:finite_particle_comparison}
\end{figure}

Figure~\ref{fig:finite_particle_comparison}(a) shows
the exit error as a function of the entrance energy
deviation $\delta$ for the nominal offsets.
In panel~(b), the offsets are scaled according to
\begin{equation}
    (a,b)=\lambda(a_0,b_0),
\end{equation}
at a fixed energy deviation of $\delta=1\%$.

Despite using only a first-order transfer matrix,
the analytical model gives smaller errors than the
second-order bend-slice representation over the
parameter ranges considered.

An accurate reference trajectory is also important
for the global survey of the beamline. The survey
determines the position and orientation of the
reference frame for subsequent elements.
For the analytical model, it is obtained by integrating
the reference motion and Bishop-frame equations using
the varying curvatures of Eq.~\eqref{eq:lin_reference},
as described in Appendix~\ref{app:survey}.
For the bend-slice model, we use the standard
Ocelot bend survey, following the conventional MAD8 bend geometry. Both are compared with the
reference trajectory obtained from direct integration
of the Lorentz equations.

\begin{figure}[!htbp]
    \centering
    \includegraphics[width=\textwidth]
    {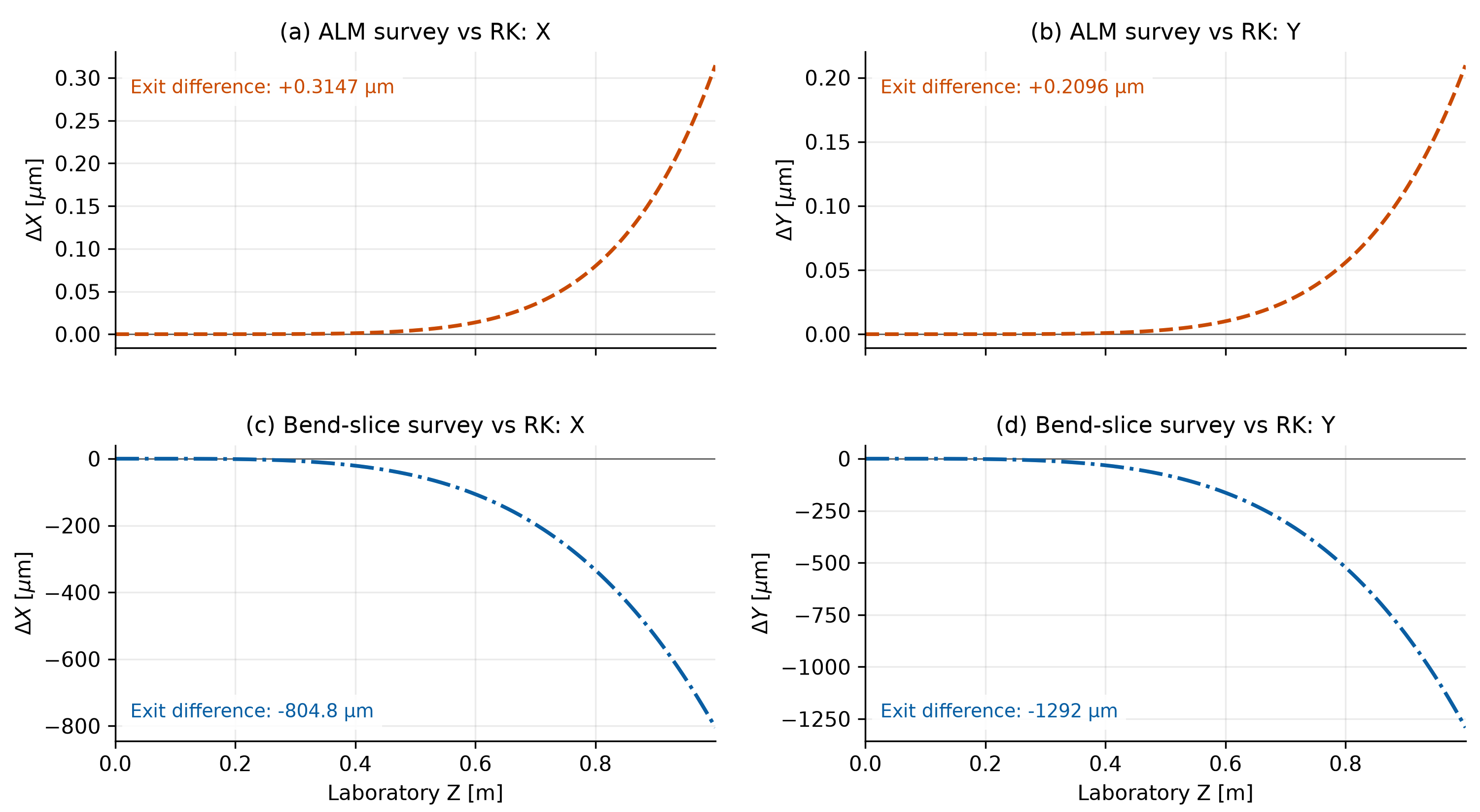}
    \caption{
    Reference-survey position differences relative to
    direct Lorentz tracking.
    (a,b) Horizontal and vertical residuals of the
    analytical-model survey.
    (c,d) Corresponding residuals of the bend-slice
    survey with 8192 bend pairs.
    All positions are compared at the same laboratory
    longitudinal coordinates $Z$.
    }
    \label{fig:survey_comparison}
\end{figure}

Figure~\ref{fig:survey_comparison} shows the reference
position differences evaluated at the same laboratory
longitudinal coordinates. The analytical-model survey
agrees closely with numerical tracking, while the
bend-slice survey exhibits much larger deviations,
reaching the millimetre scale.

This difference does not vanish with increasing slice
number. Since the bend curvatures are fixed by the
entrance offsets, the sliced representation converges
to a constant-curvature geometry rather than to the
actual quadrupole reference trajectory. The analytical
survey, in contrast, accounts for the longitudinal
variation of both curvature components.

\FloatBarrier
\section{Conclusion}

We have developed a compact analytical linear model for
beam transport through a quadrupole with offsets in both
transverse planes. Within the weak-curvature approximation,
the model provides a symplectic $6\times6$ transfer matrix
that accounts for the longitudinal variation of the
reference curvature and the resulting dispersion in
both planes.

Comparison with direct numerical tracking shows that
the analytical model reproduces the dispersive transport
more accurately than the original bend-slice representation,
despite using only a first-order map. The improvement is
also evident in the reference survey, where the
fixed-curvature bend-slice model exhibits systematic
errors that cannot be eliminated by increasing the
number of slices.

The model has been implemented in \textsc{Ocelot}
as a new lattice element, \texttt{XYQuadrupole},
providing both the linear transfer matrix and reference
geometry without replacing a single quadrupole by
a large number of bend elements.

\begin{acknowledgments}
This work was supported by operation funds of the European XFEL.

The author thanks Nina Golubeva and Winni Decking for developing the original sliced-bend representation used at the European XFEL. The author also thanks Nina Golubeva and Evgeniy Schneidmiller for valuable discussions.

Special thanks are due to Igor Zagorodnov for his encouragement to bring this work, initiated more than a year ago, to completion and for many fruitful discussions.

The author used ChatGPT (OpenAI) during the preparation of this manuscript to assist with language editing and to improve clarity and structure.
\end{acknowledgments}

\appendix

\section{Construction of the local vector potential}
\label{app:gauge}

The analytical linear model of the main text uses only those terms of the
local vector potential that contribute to the quadratic Hamiltonian. Here we
first construct the local vector potential directly from the ideal
quadrupole body field and then show explicitly how the simplified potential
used in the main text follows under the approximations of the analytical
model.

\subsection{Vector potential in the moving Bishop frame}

In the fixed magnet coordinates \((X,Y,Z)\), the normalized magnetic field is
\begin{equation}
\bb
=
k_1
\begin{pmatrix}
Y\\
X\\
0
\end{pmatrix}.
\label{eq:app_labfield}
\end{equation}
A convenient laboratory vector potential satisfying
\(\bb=\nabla\times\mathbf a\) is
\begin{equation}
\mathbf a^{\rm lab}
=
a_Z\,\mathbf E_Z,
\qquad
a_Z=
\frac{k_1}{2}\left(Y^2-X^2\right).
\label{eq:app_labpotential}
\end{equation}

The local coordinates are related to the laboratory coordinates by
\begin{equation}
\br(x,y,s)
=
\br_0(s)
+x\bx(s)
+y\by(s),
\label{eq:app_local_position}
\end{equation}
where \((\bx,\by,\bt)\) is the Bishop frame introduced in
Sec.~\ref{sec:hamiltonian}. The corresponding metric factor is
\[
g=1+h_xx+h_yy .
\]

Projecting Eq.~\eqref{eq:app_labpotential} onto the moving frame gives
\begin{equation}
a_x^{\rm raw}=a_Z e_{xZ},
\qquad
a_y^{\rm raw}=a_Z e_{yZ},
\qquad
\mathcal A_s^{\rm raw}=g\,a_Zt_Z,
\label{eq:app_raw}
\end{equation}
where the coefficient along \(ds\) is defined by
\[
\mathbf a\cdot d\br
=
a_x\,dx+a_y\,dy+\mathcal A_s\,ds .
\]
Equation~\eqref{eq:app_raw} describes the same ideal quadrupole
body field as Eq.~\eqref{eq:app_labfield}; only the coordinate description
has changed.

For the Hamiltonian it is convenient to use the transverse radial
(Poincar\'e) gauge~\cite{Brittin1982},
\begin{equation}
x a_x+y a_y=0.
\label{eq:app_radial_condition}
\end{equation}
Starting from the projected potential of Eq.~\eqref{eq:app_raw}, we perform
the gauge transformation
\begin{equation}
a_x=a_x^{\rm raw}+\partial_x\chi,
\qquad
a_y=a_y^{\rm raw}+\partial_y\chi,
\qquad
\mathcal A_s=\mathcal A_s^{\rm raw}+\partial_s\chi .
\label{eq:app_gauge_transform}
\end{equation}
A gauge function satisfying Eq.~\eqref{eq:app_radial_condition} is
\begin{equation}
\chi(x,y,s)
=
-\int_0^1
\left[
x\,a_x^{\rm raw}(\lambda x,\lambda y,s)
+
y\,a_y^{\rm raw}(\lambda x,\lambda y,s)
\right]\,d\lambda ,
\label{eq:app_gauge}
\end{equation}
where \(0\le\lambda\le1\) parametrizes the straight radial segment in the
local transverse plane.

Indeed, substituting Eq.~\eqref{eq:app_gauge} into
Eq.~\eqref{eq:app_gauge_transform} gives
\[
x a_x+y a_y=0.
\]
Equivalently, the transverse components in this gauge may be written as
\begin{equation}
a_x
=
-y\int_0^1
\lambda\,b_s(\lambda x,\lambda y,s)\,d\lambda,
\qquad
a_y
=
x\int_0^1
\lambda\,b_s(\lambda x,\lambda y,s)\,d\lambda .
\label{eq:app_radial_transverse}
\end{equation}
Thus the transverse vector-potential components are determined entirely by
the longitudinal magnetic-field component \(b_s\).

\subsection{Reduction to the analytical model}

The construction above gives the local vector potential for the ideal
quadrupole body field in the moving Bishop frame. We now expand this result
about the reference trajectory and retain only those terms that contribute
to the quadratic Hamiltonian under the paraxial and weak-curvature
approximations used in the analytical model.

In the radial gauge, the transverse components of the vector potential are
determined by the longitudinal magnetic-field component through
Eq.~\eqref{eq:app_radial_transverse}. On the reference trajectory,
\begin{equation}
b_s
=
\bb(\br_0)\cdot\bt
=
k_1\left(Y_0\theta_x+X_0\theta_y\right),
\label{eq:app_bs}
\end{equation}
which is second order in the reference offsets and angles. Its transverse
variation generates transverse vector-potential terms that contribute only
beyond the quadratic Hamiltonian retained here. Consequently,
\begin{equation}
a_x=a_y=0
\label{eq:app_axay_linear}
\end{equation}
to the order required for the analytical linear model.

It remains to determine the longitudinal component
\(\mathcal A_s\). On the reference trajectory,
Eq.~\eqref{eq:curvatures} gives
\[
b_x=-h_y,
\qquad
b_y=h_x.
\]
To first order in the transverse particle coordinates, the local transverse
field is
\begin{equation}
b_x=-h_y+k_1y,
\qquad
b_y=h_x+k_1x.
\label{eq:app_linear_field}
\end{equation}
Corrections associated with the rotation of the projected quadrupole
gradient are of higher reference order and are not retained in the
analytical model.

With \(a_x=a_y=0\), the Bishop-coordinate curl equations reduce to
\begin{equation}
b_x=\frac{1}{g}\partial_y\mathcal A_s,
\qquad
b_y=-\frac{1}{g}\partial_x\mathcal A_s,
\label{eq:app_linear_curl}
\end{equation}
where
\[
g=1+h_xx+h_yy.
\]
Writing
\[
u=h_xx+h_yy,
\]
Eq.~\eqref{eq:app_linear_curl} together with
Eq.~\eqref{eq:app_linear_field} gives
\[
\partial_y\mathcal A_s
=
(1+u)(-h_y+k_1y),
\]
and
\[
\partial_x\mathcal A_s
=
-(1+u)(h_x+k_1x).
\]
The terms generated by \(u\) that are linear in the particle coordinates
and proportional to products of the reference curvatures are neglected
according to the weak-curvature condition
Eq.~\eqref{eq:weak_curvature}. Terms quadratic in the transverse particle
coordinates would generate cubic terms in \(\mathcal A_s\) and therefore do
not contribute to the quadratic Hamiltonian. Thus, at the retained order,
\begin{equation}
\partial_y\mathcal A_s=-h_y+k_1y,
\qquad
\partial_x\mathcal A_s=-h_x-k_1x.
\label{eq:app_linear_As_derivatives}
\end{equation}

Integrating Eq.~\eqref{eq:app_linear_As_derivatives} gives
\begin{equation}
a_x=a_y=0,
\qquad
\mathcal A_s
=
a_0(s)
-h_x(s)x-h_y(s)y
-\frac{k_1}{2}(x^2-y^2),
\label{eq:app_linear_potential}
\end{equation}
which is Eq.~\eqref{eq:lin_potential} of the main text. Here
\[
a_0(s)=\mathcal A_s(0,0,s)
\]
is gauge dependent and does not affect the equations of motion.

It is useful to verify explicitly which field terms are reproduced by the
truncated potential of Eq.~\eqref{eq:app_linear_potential}. Substitution
into the Bishop-coordinate curl equations gives
\begin{equation}
b_x=
\frac{-h_y+k_1y}{1+h_xx+h_yy},
\qquad
b_y=
\frac{h_x+k_1x}{1+h_xx+h_yy}.
\label{eq:app_linear_field_exact}
\end{equation}
Expanding in the transverse coordinates,
\begin{equation}
\begin{aligned}
b_x&=
-h_y+k_1y
+h_yu-k_1yu-h_yu^2
+\order_\perp(3),
\\
b_y&=
h_x+k_1x
-h_xu-k_1xu+h_xu^2
+\order_\perp(3),
\end{aligned}
\label{eq:app_linear_field_expansion}
\end{equation}
where \(\order_\perp\) denotes transverse-coordinate degree. The terms
\(h_yu\) and \(-h_xu\) contain products of the reference curvatures and are
neglected according to Eq.~\eqref{eq:weak_curvature}. The terms
\(-k_1yu\) and \(-k_1xu\) are quadratic in the transverse particle
coordinates and first affect the cubic Hamiltonian.

Equation~\eqref{eq:app_linear_potential} therefore retains the terms of
the local vector potential required for the quadratic Hamiltonian of
the analytical model. The terms omitted here can be obtained systematically
from the construction of Eqs.~\eqref{eq:app_raw}--\eqref{eq:app_gauge}.
They include the longitudinal-field contribution to the transverse vector
potential, corrections associated with the rotation of the projected
quadrupole gradient, curvature-squared terms, and higher-order transverse
dynamics.

\subsection{Explicit expression for \texorpdfstring{$R_{56}$}{R56}}
\label{app:r56}

The longitudinal coefficient introduced in
Eq.~\eqref{eq:r56_integral} can be evaluated analytically using the
reference curvatures of Eq.~\eqref{eq:lin_reference} and the dispersion
functions of Eq.~\eqref{eq:dispersion_functions}. The result may be written
as
\begin{equation}
R_{56}
=
\frac{L}{\beta_0^2\gamma_0^2}
-\frac{\mathcal F_x+\mathcal F_y}{8\beta_0^2},
\end{equation}
with
\begin{equation}
\begin{aligned}
\mathcal F_x={}&
a^2
\left[
2k_1^2LS_x^2-k_1(L-C_xS_x)
\right]
\\
&+4a\theta_{x0}k_1
\left[
S_x^2-LC_xS_x
\right]
\\
&+\theta_{x0}^2
\left[
3(L-C_xS_x)-2k_1LS_x^2
\right],
\\[1mm]
\mathcal F_y={}&
b^2
\left[
2k_1^2LS_y^2+k_1(L-C_yS_y)
\right]
\\
&+4b\theta_{y0}k_1
\left[
LC_yS_y-S_y^2
\right]
\\
&+\theta_{y0}^2
\left[
3(L-C_yS_y)+2k_1LS_y^2
\right].
\end{aligned}
\label{eq:app_r56_components}
\end{equation}
Here \(C_i\) and \(S_i\) are evaluated at \(s=L\).

The first term in \(R_{56}\) is the usual relativistic longitudinal response,
while \(\mathcal F_x\) and \(\mathcal F_y\) describe the additional
path-length response generated by the dispersive motion in the two
transverse planes.

\section{Survey and global geometry}
\label{app:survey}

The transfer matrix of Eq.~\eqref{eq:R} describes particle
motion relative to the curved reference trajectory.
For lattice modelling, we also need the position and
orientation of the reference frame at the magnet exit,
which determine the geometry of subsequent elements.

Following the survey convention of MAD-8~\cite{MAD8},
the geometry of an element is described by a displacement
$\mathbf d$ and a rotation matrix $\mathcal Q$ relative
to its entrance frame:
\begin{equation}
    \br_{\rm out}=\br_{\rm in}+W_{\rm in}\mathbf d,
    \qquad
    W_{\rm out}=W_{\rm in}\mathcal Q,
    \label{eq:survey_composition}
\end{equation}
where
\begin{equation}
    W(s)=
    \begin{pmatrix}
        \bx(s)&\by(s)&\bt(s)
    \end{pmatrix}
    \label{eq:survey_frame}
\end{equation}
contains the Bishop-frame basis vectors.

\paragraph{Analytical approximation.}

For illustration, consider a reference particle entering
parallel to the magnet axis with offsets $(a,b)$.
In the paraxial approximation, its trajectory and
direction are given by Eqs.~\eqref{eq:reference_orbit}
and~\eqref{eq:reference_angles}.

For a magnet with physical longitudinal length $L_Z$,
the leading transverse displacement relative to the
entrance frame is
\begin{equation}
\begin{aligned}
    d_x &\simeq a(C_x-1),\\
    d_y &\simeq b(C_y-1),\\
    d_z &= L_Z,
\end{aligned}
\label{eq:survey_analytical_displacement}
\end{equation}
where $C_i$ and $S_i$ are evaluated at $L_Z$.

The exit direction is determined by
\begin{equation}
    \alpha=-k_1aS_x,
    \qquad
    \beta=k_1bS_y.
\end{equation}
To first order in the reference angles, the corresponding
rotation matrix is
\begin{equation}
    \mathcal Q\simeq
    \begin{pmatrix}
        1&0&\alpha\\
        0&1&\beta\\
        -\alpha&-\beta&1
    \end{pmatrix}.
    \label{eq:survey_analytical_rotation}
\end{equation}

The reference path length differs from the mechanical
length because of the transverse motion:
\begin{equation}
    L\simeq L_Z+
    \frac{1}{2}\int_0^{L_Z}
    \left(\theta_x^2+\theta_y^2\right)\,ds.
\end{equation}
Using Eq.~\eqref{eq:reference_angles}, we obtain
\begin{equation}
    L\simeq L_Z+\frac{k_1}{4}
    \left[
        a^2(L_Z-C_xS_x)
        +b^2(C_yS_y-L_Z)
    \right].
    \label{eq:survey_analytical_length}
\end{equation}

These expressions provide a simple approximation
for the reference displacement, orientation, and
path length. In particular, they show explicitly
how the survey depends on the quadrupole strength
and the entrance offsets.

\paragraph{Numerical survey in Ocelot.}

In the \texttt{XYQuadrupole} implementation, the survey
is calculated by numerical integration of the reference
trajectory together with the Bishop-frame equations,
rather than using the truncated expressions above:
\begin{equation}
\begin{aligned}
    \br_0' &= \bt,\\
    \bx' &= h_x(s)\bt,\\
    \by' &= h_y(s)\bt,\\
    \bt' &= -h_x(s)\bx-h_y(s)\by.
\end{aligned}
\label{eq:survey}
\end{equation}
The curvature functions $h_x(s)$ and $h_y(s)$ are
taken from the analytical approximation
of Eq.~\eqref{eq:lin_reference}.
The integration retains the full three-dimensional Bishop-frame evolution for the prescribed curvature functions, which remain within the paraxial approximation.

The entrance conditions are
\begin{equation}
    \br_0(0)=
    \begin{pmatrix}a\\b\\0\end{pmatrix},
    \qquad
    \bt(0)=
    \begin{pmatrix}
        \theta_{x0}\\
        \theta_{y0}\\
        \sqrt{1-\theta_{x0}^2-\theta_{y0}^2}
    \end{pmatrix}.
    \label{eq:survey_initial}
\end{equation}
For parallel entrance, $W(0)=I$. For nonzero entrance
angles, the initial frame is obtained by the smallest
rotation aligning the magnet axis $\mathbf E_Z$ with
$\bt(0)$. Using Rodrigues' formula,
\begin{equation}
    W(0)=I+[\mathbf v]_\times+
    \frac{[\mathbf v]_\times^2}{1+c},
    \label{eq:survey_R0}
\end{equation}
where
$\mathbf v=\mathbf E_Z\times\bt(0)$,
$c=\mathbf E_Z\cdot\bt(0)$, and $[\mathbf v]_\times$
denotes the cross-product matrix.

Integration to the prescribed reference arc length $L$
gives the exit position $\br_{\rm out}=\br_0(L)$
and orientation $W_{\rm out}=W(L)$.
The corresponding survey quantities are
\begin{equation}
\boxed{
    \mathbf d=W_{\rm in}^{T}
    (\br_{\rm out}-\br_{\rm in}),
    \qquad
    \mathcal Q=W_{\rm in}^{T}W_{\rm out}.
}
\label{eq:survey_output}
\end{equation}
The laboratory longitudinal advance $Z(L)$ is obtained
from the integrated trajectory. If the magnet is
specified instead by its mechanical length $L_Z$,
the integration is stopped at $Z=L_Z$, which also
determines the corresponding reference path length.

Finally, the survey allows the local particle
coordinates obtained from the transfer matrix to be
converted into laboratory coordinates. At the
reference-normal exit plane,
\begin{equation}
    \br_{\rm particle,out}
    =
    \br_{\rm out}
    +x_{\rm out}\bx(L)
    +y_{\rm out}\by(L).
    \label{eq:survey_particle_position}
\end{equation}
This plane generally differs from the mechanical
exit face of the magnet; conversion to the latter
requires an additional geometrical transformation.

\end{document}